\documentclass{llncs}

\usepackage{hyperref} % links in documents

\newcommand{\codefont}{\tt\small}
\newcommand{\code}[1]{\mbox{\codefont{#1}}}
\newcommand{\ccode}[1]{``\code{#1}''}

\usepackage{listings}
\lstnewenvironment{curry}[1][]
  {\lstset{#1}}{}
\newcommand{\listline}{\vrule width0pt depth1.5ex}

\newcommand{\incurry}[1]{\lstinline{#1}}

\begin{document}

\title{Functional Logic Program Transformations}

\author{
Michael Hanus\inst{1}%\orcidID{0000-0002-4953-8202}
\and
Steven Libby\inst{2}
}
\institute{
Institut f\"ur Informatik, Kiel University, Kiel, Germany\\
\email{mh@informatik.uni-kiel.de}
\and
University of Portland, Portland, Oregon, U.S.A.\\
\email{libbys@up.edu}
}

\maketitle

\begin{abstract}
Many tools used to process programs, like compilers, analyzers,
or verifiers, perform transformations on their intermediate
program representation, like abstract syntax trees.
Implementing such program transformations is a non-trivial task,
since it is necessary to iterate over the complete syntax tree
and apply various transformations at nodes in a tree.
In this paper we show how the features of functional logic programming
are useful to implement program transformations in a compact
and comprehensible manner.
For this purpose, we propose to write program transformations
as partially defined and non-deterministic operations.
Since the implementation of non-determinism usually
causes some overhead compared to deterministically defined operations,
we compare our approach to a deterministic transformation method.
We evaluate these alternatives for the functional logic language Curry
and its intermediate representation FlatCurry which is used in various
analysis and verification tools and compilers.
\end{abstract}

%%%%%%%%%%%%%%%%%%%%%%%%%%%%%%%%%%%%%%%%%%%%%%%%%%%%%%%%%%%%%%%%%%%%%%%%%%%%%%
\section{Introduction}
\label{sec:intro}

Program transformation is a very old idea in computer science,
dating all the way back to McCarthy \cite{McCarthy61}.
It has proved to be an important tool in program
analysis, verification, and compilation.
Functional and logic languages in particular have seen many examples
of harnessing program transformation to perform complex tasks.
Peyton Jones demonstrated how a Haskell compiler
could be constructed as a sequence of
program transformations \cite{PeytonJones96}
and how optimizations could be implemented as more program transformations.
\cite{PeytonJones97}.
Appel provided several examples of program transformations to construct an ML
compiler \cite{Appel07} including transforming the AST into
CPS form and several optimizations.
Flanagan et al's original formulation of A-normal form was given
as a set of rewrite rules \cite{Flanagan93},
and Peyton Jones et al. showed how general rewriting can be useful for
easily implementing optimizations \cite{PeytonJones01}.
Many optimizations are presented as simple program transformations.
For example, Gill's shortcut deforestation \cite{Gill93},
inlining and beta reduction \cite{PeytonJones02},
as well as more complex optimizations like partial evaluation
\cite{Jones93}.

Functional logic languages such as
Curry\footnote{\url{https://www.curry-lang.org/}}
also take advantage of program transformations.
The RICE compiler is implemented as
a series of program transformations \cite{Libby23}.
Hanus also showed how program transformations can 
be used to simplify the verification of Curry programs \cite{Hanus26SCICO}.
Peem{\"o}ller's partial evaluator also uses several program transformations
\cite{Peemoller14}.

Although the theory of transformations is usually presented
as a simple and elegant way to work with complex structures,
the reality is usually much more complex.
For example, the A-normal from transformation \cite{Flanagan93}
is given as a set of three rewrite rules with an evaluation context.
However, the actual implementation is a half a page of Scheme code
using a moderately complex function
to build up a continuation to perform the rewrite.
While it is remarkable that the implementation is only a half page,
it has lost the readability of the original rewrite rules.
This is a reoccurring theme in program analysis and compilers.
The program transformations in the GHC compiler
are often long inscrutable functions that bear little resemblance
to the transformation they implement.

In this paper, we show how features from
functional logic programming can be used
to represent transformations in a modular, composable way.
The sections that follow review functional logic programming and Curry,
describe its intermediate representation called FlatCurry,
introduce our approach to defining transformations,
examine their results, and conclude.

%%%%%%%%%%%%%%%%%%%%%%%%%%%%%%%%%%%%%%%%%%%%%%%%%%%%%%%%%%%%%%%%%%%%%%%%%%%%%%
\section{Functional Logic Programming with Curry}
\label{sec:basics}

The declarative language Curry \cite{Hanus16Curry} supports features
from functional programming (demand-driven evaluation, strong typing,
higher-order functions) as well as from logic programming
(computing with partial information, unification, constraints),
see \cite{AntoyHanus10CACM,Hanus13} for surveys.
The syntax of Curry is close to Haskell \cite{PeytonJones03Haskell}.
In addition to Haskell, Curry applies rules
with overlapping left-hand sides in a (don't know) non-deterministic manner
(where Haskell always selects the first matching rule)
and allows \emph{free} (\emph{logic}) \emph{variables} in conditions
and right-hand sides of defining rules.
The operational semantics is based on demand-driven evaluation
which is optimal for large classes of programs
\cite{AntoyEchahedHanus00JACM}.

Similarly to Haskell, Curry is strongly typed so that a program
consists of data type definitions defining the \emph{constructors}
of these types and
\emph{functions} or \emph{operations} on these types.
The following simple example defines
the concatenation operation \ccode{++} on lists
and an operation \code{adjDup} which returns an adjacent duplicate
number in a list:\footnote{To check some unintended errors,
Curry requires the explicit declaration
of free variables, as \code{x} in the rule of \code{adjDup}.
This is not necessary for anonymous variables which are
denoted by an underscore.}
\begin{curry}
(++) :: [a] -> [a] -> [a]      adjDup :: [Int] -> Int
[]     ++ ys = ys                adjDup xs | xs =:= _$\,$++$\,$[x,x]$\,$++$\,$_
(x:xs) ++ ys = x : (xs ++ ys)              = x    where$\;$x$\;$free
\end{curry}
\code{adjDup} exploits the unification operator \ccode{=:=}
which instantiates free variables when both expressions
are evaluated and unified.
\code{adjDup} is also called a \emph{non-deterministic operation}
since it might deliver more than one result for a fixed argument,
e.g., \code{adjDup$\,$[1,2,2,1,3,3,4]} yields \code{2} and \code{3}.
Non-deterministic operations, which are interpreted as mappings
from values into sets of values \cite{GonzalezEtAl99},
are an important feature of contemporary functional logic languages.
One particularly important operation is the choice operator \ccode{?}
which non-deterministically returns one of its
two arguments.
As we will see, such operations are useful to specify
program transformations in a compact manner.

The operation \code{adjDup} is reducible if the actual argument
has the form as specified in the right-hand side of the
condition's equation.
For such cases, Curry supports a more compact notation:
\begin{curry}
adjDup' :: [Int] -> Int
adjDup' (_$\,$++$\,$[x,x]$\,$++$\,$_) = x
\end{curry}
Since the pattern used in \code{adjDup'} contains the defined function
\ccode{++}, it is called a \emph{functional pattern}.
A functional pattern denotes all standard patterns to which the
functional pattern can be evaluated.
Functional pattern matching can be efficiently implemented
by a specific unification procedure \cite{AntoyHanus05LOPSTR}.
Functional patterns can express pattern matching
at arbitrary depths so that they are useful
to specify program transformations.

Functional patterns can be supported in Curry due to its
logic programming features, i.e., the ability to deal
with non-deterministic and failing computations.
%For the latter, there is also a predefined operation
%\code{failed} which has no value.
To control non-deterministic computations and failures,
Curry supports \emph{encapsulated search operators}
which return all or some values of an expression.
In this paper, we use only the search operator
\begin{curry}
oneValue :: a -> Maybe a
\end{curry}
which returns \code{Nothing} if the argument has no value,
otherwise \code{Just} some value.\footnote{The precise selection
of the value is unspecified so that the operation is considered
unsafe (there are also other more complex declarative encapsulation
operators in Curry). Since we are not interested in the confluence
of program transformations, this slightly non-declarative behavior
is acceptable.}

%%%%%%%%%%%%%%%%%%%%%%%%%%%%%%%%%%%%%%%%%%%%%%%%%%%%%%%%%%%%%%%%%%%%%%%%%%%%%%
\section{FlatCurry: An Intermediate Represention for Curry Programs}
\label{sec:flatcurry}

Curry has many more features than described in the previous section,
like type classes, monadic I/O, modules, etc.
To avoid the consideration of all these features
in language processing tools for Curry (compilers, analyzers,\ldots),
such tools often use an intermediate language where the
syntactic sugar of the source language has been eliminated
and the pattern matching strategy is explicit.
This intermediate language is called FlatCurry and will be the
basis of example program transformations described in this paper
so that we describe it in more detail.
Apart from compilers, FlatCurry has been used
to specify the operational semantics
of Curry programs \cite{AlbertHanusHuchOliverVidal05},
to implement a modular framework for the analysis of
Curry programs \cite{HanusSkrlac14}, or to
verify non-failing properties of Curry programs \cite{Hanus26SCICO}.

\begin{figure*}[t]
\begin{displaymath}
\begin{array}{lcl@{\hspace{5ex}}l}
P & ::= & D_1 \ldots D_m  & \mbox{(program)} \\
D & ::= & f(x_1,\ldots,x_n) = e  & \mbox{(function definition)} \\
e & ::= & x & \mbox{(variable) } \\
%  & | & c(x_1,\ldots,x_n) & \mbox{(constructor application) } \\
  & | & f(e_1,\ldots,e_n)  & \mbox{(function/constructor application) } \\
  & | & e_1~\mathit{or}~e_2 & \mbox{(disjunction) } \\
  & | & \mathit{let}~x_1,\ldots,x_n ~\mathit{free~in}~ e
       & \mbox{(free variables) } \\
  & | & \mathit{let}~x = e ~\mathit{in}~ e'
       & \mbox{(let binding) } \\
  & | & \mathit{case}~e~\mathit{of}~\{p_1\to e_1; \ldots; p_n \to e_n\}
                         & \mbox{(case expression) } \\
p & ::= & c(x_1,\ldots,x_n)     & \mbox{(pattern)} 
\end{array}
\end{displaymath}
\caption{Syntax of the intermediate language FlatCurry}\label{fig:flatcurry}
\end{figure*}

Figure~\ref{fig:flatcurry} summarizes
the abstract syntax of FlatCurry.
A FlatCurry program consists of a sequence of function definitions
(we omit data type definitions here),
where each function is defined by a single rule.
Patterns in source programs are compiled into case expressions and
overlapping rules are joined by explicit disjunctions.
The patterns in each case expression are required to be non-overlapping.

Any Curry program can be transformed into this format
\cite{Antoy01PPDP,AntoyHanusJostLibby20}.
In particular, the front end of a Curry compiler transforms source programs
into FlatCurry programs so that FlatCurry is the intermediate
language of many Curry compilers.
Therefore, it is a reasonable target for program transformations
that simplify or optimize Curry programs.

For instance, consider the following operation to insert
an element at an arbitrary position into a list:\label{ex:insert}
\begin{curry}
insert :: a -> [a] -> [a]
insert x ys     = x : ys
insert x (y:ys) = y : insert x ys
\end{curry}
This definition, which has overlapping rules,
can be transformed into the
FlatCurry definition (where we use the standard list notation)
\begin{curry}
insert(x,xs) =    x : xs
               $\mathit{or}$ $\mathit{case}$ xs $\mathit{of}$ { y:ys ->$~$y : insert(x,ys) }
\end{curry}
In order to process FlatCurry programs inside Curry programs,
there is a Curry package
\code{flatcurry}\footnote{\url{https://cpm.curry-lang.org/pkgs/flatcurry.html}}
defining data types to represent FlatCurry programs
and operations to read Curry programs and returning
the equivalent FlatCurry program as terms of these data types.
To understand the transformation examples discussed later,
we show the data types to represent FlatCurry expressions.
In order to distinguish the different kinds of applications,
the following type is used:
\begin{curry}
data CombType = FuncCall
              | ConsCall
              | FuncPartCall Int
              | ConsPartCall Int
\end{curry}
Hence, \code{FuncCall} and \code{ConsCall} are used
in applications of functions and constructors, respectively,
whereas \code{FuncPartCall} and \code{ConsPartCall} are used
in partial applications where the integer argument specifies
the number of missing arguments.
Then FlatCurry expressions are represented by the following types
(for the sake of readability, this definition is slightly simplified
compared to the actual \code{flatcurry} package):
\begin{curry}
data Expr = Var  Int
          | Comb CombType String [Expr]
          | Or   Expr Expr
          | Free [Int] Expr
          | Let  (Int, Expr) Expr
          | Case Expr [BranchExpr]$\listline$
data BranchExpr = Branch Pattern Expr$\listline$
data Pattern = Pattern String [Int]
\end{curry}
Note that variables are represented as unique integers.
For instance, consider the Boolean negation operation \code{not}.
Its FlatCurry definition is
\begin{curry}
not(x) = $\mathit{case}$ x $\mathit{of}$ { False ->$~$True ; True ->$~$False }
\end{curry}
Its right-hand side expression is represented by the following
data term (where variable $x$ has the index \code{0})::
\begin{curry}
Case (Var 0)
     [Branch (Pattern "False" []) (Comb ConsCall "True" []),
      Branch (Pattern "True"  []) (Comb ConsCall "False" [])]
\end{curry}
The prelude operation \ccode{\$}, defined by
\begin{curry}
($\$$) :: (a -> b) -> a -> b
f $\$$ x = f x
\end{curry}
is an infix application operator often used to write applications
without parentheses. The Curry expression \ccode{not \$ not True}
is represented as the following data term:
\begin{curry}
Comb FuncCall "$\$$" [Comb (FuncPartCall 1) "not" [],
                   Comb FuncCall "not" [Comb ConsCall "True []]]
\end{curry}
In the next section we discuss a program transformation to
simplify this expression by removing the call to \ccode{\$}.

%%%%%%%%%%%%%%%%%%%%%%%%%%%%%%%%%%%%%%%%%%%%%%%%%%%%%%%%%%%%%%%%%%%%%%%%%%%%%%
\section{Transformations on FlatCurry Programs}
\label{sec:transflatcurry}

Now that we have a structure to transform Curry programs,
we can define methods to transform them.

\subsection{Functional logic transformations}
\label{sec:fl-transformations}

In order to simplify the development of program transformations,
we would like a simple, composable way to represent a single transformation.
With this representation, we can separate the logic
for traversing a FlatCurry expression from the transformation itself.
The most natural option is to represent a transformation as a function
on FlatCurry expressions.
We allow our transformations to be both non-deterministic and partial.
For instance, we want to implement a transformation which moves
a local let binding out of a choice.
This can be expressed by the following transformation rules:
\[
\begin{array}{r@{~~\Rightarrow~~}l}
(\mathit{let}~x = e~\mathit{in}~e_1)~?~e_2 &
             \mathit{let}~x = e~\mathit{in}~(e_1~?~e_2)\\
e_1~?~(\mathit{let}~x = e~\mathit{in}~e_2) &
             \mathit{let}~x = e~\mathit{in}~(e_1~?~e_2)
\end{array}
\]
This transformation is correct under the assumption that local
variables always have distinct identifiers (which is ensured by the Curry
front end).
The immediate implementation in Curry is shown in Figure~\ref{fig:orLet}.
Note that \code{orFloat} fails on non-matching arguments
and is overlapping on the expression
\begin{curry}
(let x = 1 in x) ? (let y = 1 in y)
\end{curry}
The non-determinism is not an issue here
because both rules will apply eventually
and the order of application does not matter.
This allows us to avoid encoding unnecessary control flow information
when it is not important.

\begin{figure}[t]
\begin{curry}
orFloat (Or (Let vs e1) e2) = Let vs (Or e1 e2)
orFloat (Or e1 (Let vs e2)) = Let vs (Or e1 e2)
\end{curry}
\caption{Transformation: float a let expression out of a choice expression}
\label{fig:orLet}
\end{figure}

Non-determinism also allows us to use functional patterns
in transformation rules.
The undollar example in Figure~\ref{fig:undollar}
implements the transformation
\[
f~\$~x ~\Rightarrow~ f~x
\]
The operation \code{dollar} abbreviates a FlatCurry expression
which is an application of \ccode{\$} where the first argument
is a partially applied function (and not a variable or another
expression).
The use of \code{dollar} as a pattern in \code{unDollar}
shows how functional patterns
can be used to improve the readability of transformations.
While we could have specified the entire pattern for the application
of the \ccode{\$} function in the definition of \code{unDollar},
the functional pattern makes the transformation clearer.

\begin{figure}[t]
\begin{curry}
unDollar (dollar f args miss x) 
 | miss == 1 = Comb FuncCall f (args++[x])
 | miss > 1  = Comb (FuncPartCall (miss-1)) f (args++[x])$\listline$
dollar f args miss x = Comb FuncCall "$\$$" 
                            [Comb (FuncPartCall miss) f args, x]
\end{curry}
\caption{Transformaton: remove a call to \ccode{\$} in an expression}
\label{fig:undollar}
\end{figure}

We also allow transformations to be partial functions.
This means that a transformation may not apply in all cases.
This is not an issue.
If the transformation fails to apply,
we ignore it and move to the next one.
This is particularly helpful when we
need to search for a specific subexpression
such as in the case canceling example in Figure~\ref{fig:caseCancel}
which implements the following transformation:
\[
\textit{case}~C~\textit{of}~\{\ldots;C \rightarrow e;\ldots\} ~\Rightarrow~ e
\]
We use a functional pattern to find the specific branch
that contains the value of the scrutinee of the case.
Requiring this function to be total would
necessitate searching through the cases manually until we find one
or signal a failure if one is not found.

\begin{figure}[t]
\begin{curry}
caseCancel (Case (Comb ConsCall c []) (withBranch c e)) = e$\listline$
withBranch c e = (_ ++ [Branch (Pattern c []) e] ++ _)
\end{curry}
    \caption{Case canceling transformation 
             for constructors with no arguments.}
\label{fig:caseCancel}
\end{figure}

A more subtle use of partial transformations occurs in the undollar 
example in Figure~\ref{fig:undollar}.
If the function \incurry{f} is a partial application with \incurry{0}
arguments missing, then we should not turn it into a partial call
expecting \incurry{-1} arguments.
By omitting this case, we have neatly sidestepped this issue.

So far, our transformations have the type \code{Expr$\;\to\;$Expr}.
While this is sufficient for simple transformations,
it is helpful to augment this type with extra information
used by more complex transformations.
Thus, we add a further argument consisting of the index of
the next fresh variable available for use and the path
to the current subexpression we are transforming
(expressed by the type \code{Path} which is a type synonym
for a list of integers).
This allows us to generate new variables as needed,
and determine useful information such as whether
we are at the root of right-hand side expression.

%Our transformations also return
%an \code{Expr} as well as the name of the transformation,
%and the number of fresh variables used.
%The name is used for debugging,
In addition to the transformed expression,
a transformation also returns the number of fresh variables
used in the transformed expression.
This allows us to keep track of the next fresh variable efficiently.
Thus, the full type of a general expression transformation is
\begin{curry}
type ExprTransformation = (Int, Path) -> Expr -> (Expr,Int)
\end{curry}
We can lift a simple transformation of type \code{Expr$\;\to\;$Expr}
to \incurry{ExprTransformation} with the function
\begin{curry}
makeT :: (Expr -> Expr) -> ExprTransformation
makeT f = \_ e -> (f e, 0)
\end{curry}
Finally, our transformation library provides an operation
(its implementation is discussed in Sect.~\ref{sec:strategies})
\begin{curry}
transformExpr :: (() -> ExprTransformation) -> Expr -> Expr
\end{curry}
which applies a transformation repeatedly until no more transformations
can be applied.
To avoid committing to a specific choice of a non-deterministic transformation
too early, the transformation is not passed as a constant
but as a function which takes a dummy unit argument of type \incurry{()}.

Individual transformations such as \code{unDollar} or \code{orFloat}
can be useful, but the real power of this system comes from composability.
We support two types of composition.
Serial composition is applying one transform after another,
while parallel composition applies both transforms at the same time.
In our approach, applying transformations in series is just
function composition: for instance, applying first transformation \code{t1}
and afterwards transformation \code{t2} can be expressed in Curry
with the function composition operator \ccode{.} by
\begin{curry}
transformExpr (\() -> t2) . transformExpr (\() -> t1)
\end{curry}
The parallel composition of  \code{t1} and \code{t2}, i.e.,
applying both transformations \code{t1} and \code{t2}
whenever possible, can be expressed by
\begin{curry}
transformExpr (\() -> t1 ? t2)
\end{curry}
This composition is expressive
and gives us control over how we execute our transformations.
For example, we may want to float all of the let expressions out of a choice
before we remove all \ccode{\$} applications and cancel simple cases.
This can be expressed by
\begin{curry}
transformExpr (\() -> makeT unDollar ? makeT caseCancel) . 
  transformExpr (\() -> makeT orFloat)
\end{curry}

\subsection{Purely functional transformations}

Separation of concerns and composability are powerful tools in our system.
However, we may wish to avoid all non-determinism in a program transformation.
Although this sacrifices convenience,
we can still support composition
with total deterministic functions.

Deterministic transformations must be totally defined so that they have type
\incurry{Expr -> Maybe Expr}.
A successful transformation returns \code{Just e},
where a failing transformation returns \code{Nothing}.

The \code{caseCancelDET} example below shows the changes we need to
make to define a deterministic transformation.
Functional patterns are replaced with total case expressions.
However, because we need to find the branch in a list,
we need to write an auxiliary function to find the correct branch.
The case expressions ensure that pattern matching is done in sequence
so we do not have overlapping patterns.
\begin{curry}
caseCancelDET e = case e of
                    Case (Comb ConsCall c []) bs -> find c bs
                    _                            -> Nothing$\listline$
 where find _ [] = Nothing
       find c (Branch (Pattern p vs) e : bs)
         | c == p && null vs = Just e
         | otherwise         = find n bs
\end{curry}
To apply a deterministic transformation repeatedly to all subexpressions
until no more transformation is possible,
our library provides an operation
\begin{curry}
transformExprDet :: ExprTransformation -> Expr -> Expr
\end{curry}
Note that the unit argument used in \code{transformExpr}
is not necessary since the transformation argument
is always deterministic.

Sequential composition is unchanged from the previous version,
but we need a new operator for parallel composition.
The \incurry{<?>} operator tries one transformation,
and falls back on the second if the first transformation fails.
\begin{curry}
t1 <?> t2 = \env e -> case t1 env e of
                         Nothing -> t2 env e
                         answer  -> answer
\end{curry}
With this new composition operator, we can define a deterministic
version of our previous transformation with the following
(where we omit the definition of the deterministic versions
of \code{unDollear} and \code{orFloat}):
\begin{curry}
transformExprDet (makeT unDollarDet <?> makeT caseCancelDet) . 
  transformExprDet (makeT orFloatDet)
\end{curry}

\subsection{Transformation strategies}
\label{sec:strategies}

In order to apply the transformation rules shown so far
to arbitrary subexpressions of an expression (typically, the
right-hand side of a function definition), we need a transformation engine.
This is the job of the operation \code{transformExpr}.
To evaluate our framework, we use three different
strategies for applying transformations which are sketched in the following.

The first is the \emph{chaotic strategy} (\emph{CS}).
This strategy non-deterministically selects a subexpression,
tries to apply a transformation on this subexpression,
and, if possible, replaces it by the result of the transformation
This is similar to the deep selection pattern \cite{Antoy11}.
When applying a transformation rule to the selected subexpression,
the operation \code{oneValue} is used to check whether a transformation
is applicable and to ensure that only one replacement is performed.
This is the easiest way to implement
but might be inefficient for large programs.\footnote{%
An implementation of this strategy in Curry
is shown in Appendix~\ref{sec:impl-chaotic}.}

The second strategy is the \emph{deterministic strategy} (\emph{DS}).
This strategy traverses the expression in a bottom-up fashion,
while attempting to apply a purely functional transformation
at each subexpression.
If it applies, the subexpression is replaced and we try again,
otherwise we move on.

The final strategy is a combination of the two
that we call the \emph{mixed strategy} (\code{MS}).
This strategy still traverses the syntax tree in a bottom up manner,
but non-deterministic transformation rules are applied at each step
where \code{oneValue} is used to force the result to be deterministic.
This improves on the chaotic strategy because we do not need to recompute the
path every time a transformation applies.

With a framework for creating, composing, and applying transformations,
we turn our attention to evaluating 
the effectiveness of the different strategies.
We assess both the performance as well
as the development cycle for transformations.

%%%%%%%%%%%%%%%%%%%%%%%%%%%%%%%%%%%%%%%%%%%%%%%%%%%%%%%%%%%%%%%%%%%%%%%%%%%%%%
\section{Benchmarks}

We have presented different methods to implement transformations
in a declarative language.
Functional logic transformations exploit partially defined
non-deterministic operations to specify transformations
in a comprehensible manner and avoid superfluous code
for non-matching cases.
Their actual implementation demands for logic programming features
like controlling failures and non-determinism via encapsulated search.
In contrast, purely functional transformations require
the encoding of all cases in a deterministic manner
but their implementation does not require logic programming features.
Since the use of logic features requires some additional efforts
at run time, it is interesting to know about the price to pay
for supporting compact and comprehensible transformation specifications.
Therefore, we evaluate the approaches discussed above in this section.

% NOTE: The benchmark numbers are obtained by executing
%
% fcy-trans Prelude Data.Char Data.Either Data.List Data.Maybe
%           Numeric System.Console.GetOpt System.IO --statistics
%           -r bench
%
\begin{table}[t]
\centering
\begin{tabular}{l@{~}|@{~}rrr@{~~}|rrr|rrr}
\hline
 & & & & \multicolumn{3}{c}{PAKCS} & \multicolumn{3}{|c}{KiCS2} \\
Module & Size & Funcs & Trans & ~~CS & ~~~MS & ~~DS & ~~CS & ~~MS & ~~DS \\
\hline
Prelude & 72485 & 1285          & 3 & 955 & 1494 & 321 & 795 & 306 & 148 \\
Data.Char & 2190 & 9               & 0 & 22 & 49 & 7   & 18 & 7 & 4 \\    
Data.Either & 1693 & 11            & 0 & 3 & 3 & 1     & 1 & 1 & 1 \\      
Data.List & 14841 & 87             & 0 & 54 & 65 & 20  & 24 & 18 & 9 \\     
Data.Maybe & 1809 & 9              & 0 & 4 & 6 & 1     & 3 & 2 & 1 \\      
Numeric & 3494 & 7                 & 4 & 9 & 18 & 4    & 17 & 17 & 2 \\      
System.Console.GetOpt & 17328 & 47 & 0 & 66 & 119 & 21 & 38 & 23 & 11 \\   
System.IO & 6223 & 51              & 0 & 15 & 19 & 7   & 5 & 5 & 3    \\
\hline
\end{tabular}\\[1ex]
\caption{Transforming standard libraries with the transformations
of Sect.~\ref{sec:fl-transformations}}
\label{table:transformtimes}
\end{table}

We implemented the chaotic, mixed, and deterministic transformation
strategies in Curry so that the transformation rules can be written
as shown above.
In particular, transformations written in a functional logic style
can be used for both the chaotic and mixed strategy,
whereas the deterministic strategy requires the more complex
style of deterministic transformations in the form of totally defined
operations.

In the first set of benchmarks,
we applied all three transformation rules shown in
Sect.~\ref{sec:fl-transformations} to various standard libraries
contained in Curry distributions.
For each module, Table~\ref{table:transformtimes} shows
the size of the Curry source file,
the number of defined functions in the corresponding FlatCurry program
(note that the transformations are applied to the right-hand side
of each function),
the number of transformations performed in the module,
and the time (in milliseconds) to apply the various strategies
described in Sect.~\ref{sec:strategies}.\footnote{%
We measured the transformation times on a Linux machine % Laptop petrus
running Ubuntu 22.04 with an Intel Core i7-1165G7 (2.80GHz) processor
with eight cores.}
Furthermore, we executed the benchmarks with the Curry implementations
PAKCS \cite{Hanus25PAKCS}, which compiles into Prolog, and
KiCS2 \cite{BrasselHanusPeemoellerReck11}, which compiles into Haskell.
Since the standard libraries evolved over years so that
superfluous pieces of code are avoided, there are only
a few transformations which can be applied.
Thus, this benchmark mainly evaluates the time to try to apply
transformations at all positions in a program w.r.t.\ different strategies.

As one can expect, the deterministic transformation strategy
is the fastest.
However, the times with the mixed strategy are not so much worse.
Thus, there is no need to rewrite functional logic transformations
into the more complex purely deterministic style, in particular,
if an efficient Curry implementation like KiCS2 is used.
It is also interesting that the chaotic strategy
is faster than the mixed strategy when it is executed with PAKCS..
The reason is unclear but it should be noted that Prolog
has a direct support for non-determinism whereas KiCS2
explicitly implements non-determinism via search tree structures.

\begin{table}[t]
\centering
\begin{tabular}{l@{~}|@{~}rrr@{~~}|rrr|rrr}
\hline
 & & & & \multicolumn{3}{c}{PAKCS} & \multicolumn{3}{|c}{KiCS2} \\
Module & Size & Funcs & Trans & ~~CS & ~~~MS & ~~DS & ~~CS & ~~MS & ~~DS \\
\hline
              Prelude & 72485 &  1285 &  5779 & 39339 & 77830 & 3042 & 15660 & 5272 & 2740 \\
            Data.Char &  2190 &     9 &   163 &  1358 &  3189 &   90 &   365 &   86 &   76 \\
          Data.Either &  1693 &    11 &    12 &     6 &    10 &    2 &     2 &    3 &    3 \\
            Data.List & 14841 &    87 &   237 &   481 &   870 &   70 &   117 &   61 &   56 \\
           Data.Maybe &  1809 &     9 &    29 &   103 &   175 &   13 &    39 &   11 &   11 \\
              Numeric &  3494 &     7 &    47 &    54 &    93 &    9 &    10 &    8 &    6 \\
System.Console.GetOpt & 17328 &    47 &   418 &  1483 &  2566 &  139 &   331 &  126 &  108 \\
            System.IO &  6223 &    51 &    89 &    50 &    89 &   15 &     9 &   11 &    9 \\
\hline
\end{tabular}\\[1ex]
\caption{Transforming standard libraries to A-normal form}
\label{table:anf-transformtimes}
\end{table}

In order to get an impression for the transformation behavior
when many transformations can actually be applied,
we tested our implementation with a transformation
of FlatCurry programs into their A-Normal Form (ANF) \cite{Flanagan93}.
ANF requires that operations or constructors
are only applied to variables.
ANF is used in many compilers and also in the specification
of operational semantics for functional \cite{Launchbury93}
and functional logic \cite{AlbertHanusHuchOliverVidal05} programs.
The ANF transformation replaces non-variable arguments by fresh
variables which are introduced in let bindings.
For instance, the A-Normal Form of the operation \code{insert}
defined in Sect.~\ref{ex:insert} is
\begin{curry}
insert(x,xs) =    x : xs
               $\mathit{or}$ $\mathit{case}$ xs $\mathit{of}$ { y:ys ->$~\mathit{let}~$z = insert(x,ys)
                                       $\mathit{in}$ y : z }
\end{curry}
The functional logic ANF transformation basically consists of
a single rule which replaces a non-variable argument
by a new variable and the corresponding let binding.
On the other hand, the purely functional ANF transformation
has to implement the same transformation but must also
consider all other expressions in order to avoid a failure.

The results of applying the ANF transformation to the set of standard libraries
are shown in Table~\ref{table:anf-transformtimes}.
The run times indicate that the differences between the non-deterministic
and the deterministic transformation can get larger
when many transformations are applicable,
which is the case for the Prolog-based PAKCS compiler.
However, for the more efficient Haskell-based KiCS2 compiler,
the differences are not so high, in particular, when the
mixed strategy is used instead of the chaotic strategy.
Since the same transformation rules can be used
for both strategies, the mixed strategy can always be preferred.
Taking into account that functional logic transformations
are easier to implement, the differences in the absolute timings
are acceptable when an efficient Curry system, like KiCS2, is used.

%%%%%%%%%%%%%%%%%%%%%%%%%%%%%%%%%%%%%%%%%%%%%%%%%%%%%%%%%%%%%%%%%%%%%%%%%%%%%%
\section{Conclusions}
\label{sec:concl}

In this paper we presented methods to implement program
transformations in a declarative manner.
Our examples target the intermediate language FlatCurry
but similar transformations can be written for other languages
if their abstract syntax trees are represented as data terms.
Although such transformations can be implemented in
any programming language, we showed that the functional logic programming
features of Curry are useful to write compact and comprehensible
transformations so that the implemented code is
quite similar to specifications of such transformations.
The definition of transformations as
partially defined and non-deterministic operations
avoids writing superfluous code to control the transformations.

Qualitatively, our system for program transformations
offers several advantages to writing the transformations by hand.
One of the most significant benefits is that the transformations
are written entirely in Curry.
There is no new language to learn or tool to install.
Instead, each transformation is a Curry function.
This allows Curry developers to write complex transformations
without needing to learn a new system.
Developers with experience in functional languages
can typically understand a transformation with limited explanation.

Another advantage is that transformations are extensible.
Because transformations are Curry functions,
we can make them as complex as necessary.
Transformations are not restricted to being simple rewrite rules,
but can involve arbitrary computations.
Transformations can be extended to handle additional information.
For example, a beta-reduction transformation can
take a map containing information about previously seen functions.
We simply add another parameter to the transformation.

A third advantage is that our system
can easily be extended to report which transformations are applied.
By supplying a name to each transformation,
it becomes easy to reconstruct the entire transformation derivation.
This is an important tool for debugging transformations.
It enables developers to see how expressions evolved from
the original form to the final result,
which is often difficult when working with several transformations.

Our system has been used extensively in the RICE compiler \cite{Libby23}
to implement complex transformations including
conversion to A-normal form, variable inlining, and beta-reduction.
Development of these optimizations was usually straightforward,
and we were able to see how the optimizations interacted with one another.
It also made it easier to modify the optimizations to make them more effective.

Our system provides a convenient, concise method of specifying 
program transformations entirely in Curry.
By allowing both partial and non-deterministic operations,
we are able to focus on the transformations themselves.
This system is extensible
and has been proven effective in large programs like the RICE compiler.

\bibliographystyle{plain}
%\bibliography{mh}
\bibliography{paper}

\newpage
\appendix

\section{Implementation of the Chaotic Strategy}
\label{sec:impl-chaotic}

In the following we show the implementation of the chaotic transformation
strategy in Curry.
First, we define the non-deterministic operation \code{subExpOf}
which returns, for a given FlatCurry expression,
some subexpression and its path (a list of integers representing
argument positions counted from zero).
\begin{curry}
subExpOf :: Expr -> (Path,Expr)
subExpOf e = ([],e) -- the subexpression is the entire expression
subExpOf (Comb _ _ args) =
  uncurry extendPath $\$$ anyOf (zip [0..] (map subExpOf args))
subExpOf (Let (_,e) _) = extendPath 0 (subExpOf e)
subExpOf (Let _     e) = extendPath 1 (subExpOf e)
subExpOf (Free _ e) = extendPath 1 (subExpOf e)
subExpOf (Or e1 e2) =
  extendPath 0 (subExpOf e1) ? extendPath 1 (subExpOf e2)
subExpOf (Case ce _) = extendPath 0 (subExpOf ce)
subExpOf (Case _ bs) = uncurry extendPath $\$$
  anyOf (zip [1..] (map (subExpOf . branchExp) bs))
 where branchExp (Branch _ be) = be
\end{curry}
The auxiliary operation \code{extendPath}
extends the path component by one position:
\begin{curry}
extendPath :: Int -> (Path,Expr) -> (Path,Expr)
extendPath pos (p,e) = (pos:p, e)
\end{curry}
Now we can define the chaotic strategy as the operation
\code{transformExpr} which simplifies a FlatCurry expression
by iteratively applying the given transformation rule to
some subexpression.
Thus, a transformation step is implemented by selecting a subexpression
with \code{subExpOf} where a rule can be applied (implemented by
the local operation \code{tryTransExpr}).
To make the overall strategy deterministic and non-failing,
we control the transformation with \code{oneValue}.
The operation \code{newVar} returns the next fresh variable not occurring
in an expression, and the operation \code{replace}
replaces a subexpression in an expression at the given position
(its definition is a similar case distinction as in \code{subExpOf}).
\begin{curry}
transformExpr :: (() -> ExprTransformation) -> Expr -> Expr
transformExpr trans n e = runTrExpr trans (newVar e) e$\listline$
runTrExpr :: (()$\;$->$\;$ExprTransformation) -> Int -> Expr -> Expr
runTrExpr trans nvar exp =
  case oneValue (tryTransExpr nvar exp) of
    Nothing            -> exp -- no transformation applicable
    Just (p, (e',nvs)) -> runTrExpr trans (nvar+nvs)
                                     (replace exp p e')
 where
  tryTransExpr v e = let (p,se) = subExpOf e
                     in (p, trans () (v,p) se)

\end{curry}

\end{document}